\documentclass[12pt]{rspublic}
\usepackage[dvips,final]{graphicx}
\usepackage{epsfig}
\usepackage{epic}
\usepackage{eepic}
\usepackage{amssymb,amsmath}
\usepackage{times}
\usepackage{astron-jh}
\usepackage{subfigure}
\begin{document}

\hyphenation{di-men-sio-nal}
\newcommand{\vect}[1]{{ {\bf #1  }}} 
\newcommand{\uvect}[1]{{ \hat{\bf #1  }} }
\newcommand{\ci}{
	{
		{ {\bf c}}_i
	}
}
\newcommand{\fixme}[1]{

{ \bf{ ***FIXME: #1 }}

}
\newcommand{\half}{\frac{1}{2}}
\newcommand{\cop}{\Omega_i^\sigma}
\newcommand{\copbgk}{{\Omega_i^\sigma}_{\mathrm{BGK}}}
\newcommand{\Dop}{\mathbb D}
\newcommand{\tausig}{\tau^\sigma}
\newcommand{\xsig}{x^\sigma}
\newcommand{\tausigb}{\tau^\bar{\sigma}}
\newcommand{\psisig}{\psi^\sigma}
\newcommand{\psisigb}{\psi^{\bar{\sigma}}}
\newcommand{\nusig}{\nu^\sigma}
\newcommand{\msig}{m^\sigma}
\newcommand{\nsig}{n^\sigma}
\newcommand{\usig}{u^\sigma}
\newcommand{\usiga}{u^\sigma_\alpha}
\newcommand{\Fsig}{F^\sigma}
\newcommand{\Fsiga}{F^\sigma_\alpha}
\newcommand{\upr}{u'}
\newcommand{\upra}{{u'}_\alpha}
\newcommand{\vsig}{v^\sigma}
\newcommand{\vsiga}{v^\sigma_\alpha}
\newcommand{\sumsig}{\sum_\sigma}
\newcommand{\sumsigb}{\sum_{\bar{\sigma}}}
\newcommand{\sumsigsigb}{\sum_{\sigma\bar{\sigma}}}
\newcommand{\sumi}{\sum_i}
\newcommand{\msi}{\msig\sum_i}
\newcommand{\ciao}{c_{i{\alpha_1}}}
\newcommand{\cian}{c_{i{\alpha_n}}}
\newcommand{\cia}{c_{i\alpha}}
\newcommand{\cib}{c_{i\beta}}
\newcommand{\cig}{c_{i\gamma}}
\newcommand{\cid}{c_{i\delta}}
\newcommand{\cs}{c_{\mathrm{s}}}
\newcommand{\rhosig}{\rho^\sigma}
\newcommand{\frt}{\frac{\rhosig}{\tausig}}
\newcommand{\xt}{(\vect{x},t)}
\newcommand{\xpct}{(\vect{x}+\ci,t)}
\newcommand{\xpc}{(\vect{x}+\ci)}
\newcommand{\Ua}{U_\alpha}
\newcommand{\fis}{f_i^\sigma}
\newcommand{\fisb}{\bar{f}_i^\sigma}
\newcommand{\Nis}{N_i^\sigma}
\newcommand{\NiU}{N_i^\sigma({\bf U})}
\newcommand{\Niu}{N_i^\sigma({\bf u})}
\newcommand{\Nivs}{N_i^\sigma({\bf v}^\sigma)}
\newcommand{\Ti}{T_i}
\newcommand{\Ta}{{T}^{(1)}_{\alpha}}
\newcommand{\Tab}{{T}^{(2)}_{\alpha\beta}}
\newcommand{\Tabg}{{T}^{(3)}_{\alpha\beta\gamma}}
\newcommand{\kronab}{\delta_{\alpha\beta}}
\newcommand{\kronag}{\delta_{\alpha\gamma}}
\newcommand{\kronbg}{\delta_{\beta\gamma}}
\newcommand{\oz}[1]{{#1}^{(0)}}
\newcommand{\oo}[1]{{#1}^{(1)}}
\newcommand{\ot}[1]{{#1}^{(2)}}
\newcommand{\ordn}[1]{{#1}^{(n)}}
\newcommand{\partiald}[2]{
	\frac { \partial #1 } { \partial #2 }
}
\newcommand{\partialdd}[2]{
	\frac { \partial^2 #1 } { \partial {#2}^2 }
}

\newcommand{\partialop}[1]{
	\frac { \partial } { \partial #1 }
}
\newcommand{\partialopop}[1]{
	\frac { \partial^2 } { \partial {#1}^2 }
}
\newcommand{\dal}{\partial_\alpha}
\newcommand{\dbe}{\partial_\beta}
\newcommand{\dga}{\partial_\gamma}
\newcommand{\dt}{\partial_{t}}
\newcommand{\dit}{\partial_{1t}}
\newcommand{\dtt}{\partial_{2t}}
\newcommand{\Psa}{\Pi^\sigma_\alpha}
\newcommand{\Psab}{\Pi^\sigma_{\alpha\beta}}
\newcommand{\Psabg}{\Pi^\sigma_{\alpha\beta\gamma}}
\newcommand{\Pa}{\Pi_\alpha}
\newcommand{\Pab}{\Pi_{\alpha\beta}}
\newcommand{\Pabg}{\Pi_{\alpha\beta\gamma}}
\newcommand{\ep}{\epsilon}

\title{Large-scale lattice Boltzmann simulations of complex fluids: advances
through the advent of computational grids}

\author[J.~Harting, J.~Chin, M.~Venturoli, and P.V.~Coveney]{Jens Harting$^1$, Jonathan
Chin$^{2}$, Maddalena Venturoli$^{2,3}$, and Peter V. Coveney$^2$}

\affiliation{$^1$Institute for Computational Physics, University of
Stuttgart, Pfaffenwaldring 27, D-70569 Stuttgart, Germany\\ $^2$Centre for
Computational Science, Christopher Ingold Laboratories, University College
London, 20 Gordon Street, London WC1H 0AJ, UK\\ $^3$Schlumberger Cambridge
Research, High Cross, Madingley Road, Cambridge CB3 0EL, UK
}
\label{firstpage}

\maketitle

\begin{center}
\begin{abstract}{Lattice-Boltzmann, complex fluids, grid
computing, computational steering}

During the last two years the RealityGrid project has allowed us to be
one of the few {\it scientific} groups involved in the development of
computational grids. Since smoothly working production grids are not yet
available, we have been able to substantially influence the direction of
software development and grid deployment within the project.
In this paper we review our results from large scale three-dimensional
lattice Boltzmann simulations performed over the last two years. We
describe how the proactive use of computational steering and advanced
job migration and visualization techniques enabled us to do our
scientific work more efficiently. 
The projects reported on in this paper are studies of complex fluid
flows under shear or in porous media, as well as large-scale parameter
searches, and  studies of the self-organisation of liquid cubic
mesophases.

\end{abstract}
\end{center}

\section{Introduction}

In recent years there has emerged a class of fluid dynamical problems,
called ``complex fluids'',
which involve both hydrodynamic flow effects and complex interactions
between fluid particles. Computationally, such problems are too large
and expensive to tackle with atomistic methods such as molecular
dynamics, yet they require too much molecular detail for continuum
Navier-Stokes approaches.

Algorithms which work at an intermediate or ``mesoscale'' level of
description in order to solve these problems have been developed in
response, including Dissipative Particle
Dynamics\cite{bib:espanol-warren,bib:jury-bladon-cates,bib:flekkoy-coveney-defabritiis},
Lattice Gas Cellular Automata\cite{bib:rivet-boon}, the Stochastic
Rotation Dynamics of Malevanets and
Kapral\cite{bib:malevanets-kapral,bib:hashimoto-chen-ohashi,bib:sakai-chen-ohashi},
and the Lattice Boltzmann
Equation\cite{bib:succi,bib:benzi-succi-vergassola,bib:love-nekovee-coveney-chin-gonzalez-martin}.
In particular, the Lattice Boltzmann method has been found highly useful
for simulation of complex fluid flows in a wide variety of systems. This
algorithm, described in more detail below, is extremely well suited to
implementation on parallel computers, which permits very large systems
to be simulated, reaching hitherto inaccessible physical regimes. We
describe some of these calculations, and also attempts to take parallel
computing to a new scale, by coupling several supercomputers together
into a computational grid, which in turn permits easy use of techniques
such as computational steering, code migration, and real-time
visualization.

The term ``simple fluid'' usually refers to a fluid which can be described
to a good degree of approximation by macroscopic quantities only, such
as the density field $\rho(\bf{x})$, velocity field $\bf{v}(\bf{x})$,
and perhaps temperature $T(\bf{x})$. Such fluids are governed by the
well-known Navier-Stokes equations\cite{bib:faber}, which, being
nonlinear, are difficult to solve in the most general case, with the
result that numerical solution of the equations has become a common
tool for understanding the behaviour of ``simple'' fluids, such as water
or air.
Conversely, a ``complex fluid'' is one whose macroscopic flow is affected
by its microscopic properties. A good example of such a fluid is blood:
as it flows through vessels (of order millimetres wide and centimetres
long), it is subjected to shear forces, which cause red blood cells (of
order micrometres wide) to align with the flow so that they can slide
over one another more easily, causing the fluid to become less viscous;
this change in viscosity in turn affects the flow profile. Hence, the
macroscopic blood flow is affected by the microscopic alignment of its
constituent cells.  Other examples of complex fluids include
biological fluids such as milk, cell organelles and cytoplasm, as
well as polymers and liquid crystals. In all of these cases, the density
and velocity fields are insufficient to describe the fluid behaviour,
and in order to understand this behaviour, it is necessary to treat
effects which occur over a very wide range of length and time scales.
This length and time scale gap makes complex fluids even more difficult
to model than ``simple'' fluids. While numerical solutions of the
macroscopic equations are possible for many simple fluids, such a level
of description may not exist for complex fluids, yet simulation of every
single molecule involved is computationally infeasible.

In a mixture containing many different fluid components, an amphiphile
is a kind of molecule which is composed of two parts, each part being
attracted towards a different fluid component. For example, soap
molecules are amphiphiles, containing a head group which is attracted
towards water, and a tail which is attracted towards oil and grease;
analogous molecules can also be formed from polymers. If many amphiphile
molecules are collected together in solution, they can exhibit highly
varied and complicated behaviour, often assembling to form amphiphile
mesophases, which are complex fluids of significant theoretical and
industrial importance. Some of these phases have long-range order, yet
remain able to flow, and are called liquid crystal mesophases. Of
particular interest to us are those with cubic symmetry, whose
properties have been studied experimentally
\cite{bib:seddon-templer,bib:seddon-templer-2,bib:czeslik-winter} in
lipid-water mixtures\cite{bib:seddon-templer}, diblock
copolymers\cite{bib:shefelbine-vigild-etal}, and in many biological
systems\cite{bib:landh}.

Over the last decade, significant effort has been invested in
understanding complex fluids through computational mesoscale modelling
techniques. These techniques do not attempt to keep track of the state
of every single constituent element of a system, nor do they use an
entirely macroscopic description; instead, an intermediate, {\it
mesoscale} model of the fluid is developed, coarse-graining microscopic
interactions enough that they are rendered amenable to simulation and
analysis, but not so much that the important details are lost. Such
approaches include Lattice Gas
Automata\cite{bib:rivet-boon,bib:fhp,bib:rothman-keller,bib:love}, the
Lattice Boltzmann
equation\cite{bib:succi,bib:benzi-succi-vergassola,bib:mcnamara-zanetti,bib:higuera-jimenez,bib:higuera-succi-benzi,bib:shan-chen,bib:lamura-gonnella-yeomans,bib:chen-boghosian-coveney,bib:chin-coveney},
Dissipative Particle
Dynamics\cite{bib:hoogerbrugge-koelman,bib:espanol-warren,bib:jury-bladon-cates},
or the Malevanets-Kapral Real-coded Lattice
Gas\cite{bib:malevanets-kapral,bib:malevanets-yeomans,bib:hashimoto-chen-ohashi,bib:sakai-chen-ohashi}.
Recently-developed
techniques\cite{bib:garcia-bell-crutchfield-alder,bib:delgado-coveney}
which use hybrid algorithms have shown much promise.

\newcommand{\fone}{f_1(\vect{r},\vect{v},t)}

All simulations described in this paper use the lattice Boltzmann
algorithm, which is a powerful method for simulating fluid dynamics. This
is due to the ease with which boundary conditions can be imposed, and with
which the model may be extended to describe mixtures of interacting
complex fluids. Rather than tracking the state of individual atoms and
molecules, the method describes the dynamics of the single-particle
distribution function of mesoscopic fluid packets.

In a continuum description, the single-particle distribution function
$\fone$ represents the density of fluid particles with position $\vect{r}$
and velocity $\vect{v}$ at time $t$, such that the density and velocity of
the macroscopically observable fluid are given by $\rho(\vect{r},t) = \int
\fone {\mathrm d}\vect{v} $ and $\vect{u}(\vect{r},t) = \int \fone
\vect{v} {\mathrm d} \vect{v}$ respectively. In the non-interacting, long
mean free path limit, with no externally applied forces, the evolution of
this function is described by Boltzmann's equation,
\begin{equation}
\label{eq:boltzmann}
\left( \dt + \vect{v} \cdot \vect{\nabla} \right) f_1
= \Omega[f_1].
\end{equation}
While the left hand side describes changes in the distribution function
due to free particle motion, the right hand side models pairwise
collisions. This collision operator $\Omega$ is an integral expression
that is often simplified~\cite{bib:bgk} to the linear
Bhatnagar-Gross-Krook (BGK) form
\begin{equation}
\label{eq:bgk}
\Omega[f] \simeq - \frac 1 \tau \left[ f - f^{\mathrm{(eq)}} \right].
\end{equation}
The BGK collision operator describes the relaxation, at a rate controlled
by a characteristic time $\tau$, towards a Maxwell-Boltzmann equilibrium
distribution $f^{\mathrm{(eq)}}$. While this is a drastic simplification,
it can be shown that distributions governed by the Boltzmann-BGK equation
conserve mass, momentum, and energy \cite{bib:succi}, and obey a
non-equilibrium form of the Second Law of
Thermodynamics~\cite{bib:liboff}.  Moreover, it can be
shown~\cite{bib:chapman-cowling,bib:liboff} that the well-known
Navier-Stokes equations for macroscopic fluid flow are obeyed on coarse
length and time scales~\cite{bib:chapman-cowling,bib:liboff}.
In a lattice Boltzmann formulation, the single-particle distribution
function is discretized in time and space. The positions $\vect{r}$ on
which $\fone$ is defined are restricted to points $\vect{r}_i$ on a
lattice, and the velocities $\vect{v}$ are restricted to a set $\ci$
joining points on the lattice. The density of particles at lattice site
$\vect{r}$ travelling with velocity $\ci$, at timestep $t$ is given by
$f_i(\vect{r},t) = f(\vect{r},\ci,t)$, while the fluid's density and
velocity are given by $\rho(\vect{r}) = \sum_i f_i(\vect{r})$ and
$\vect{u}(\vect{r}) = \sum_i f_i(\vect{r}) \ci$.
The discretized description can be evolved in two steps: the collision
step,  where particles at each lattice site are redistributed across the
velocity vectors, and the advection, where values of the post-collisional
distribution function are propagated to adjacent lattice sites.
By combining these steps, one obtains the lattice-Boltzmann equation
(LBE)
\begin{equation}
\label{eq:lbgk}
f_i(\vect{r},t+1) - f_i(\vect{r},t)
= \Omega[f] \\
= - \frac 1 \tau \left[ f_i(\vect{r},t)
- N_i\left( \rho, \vect{u} \right)\right],
\end{equation}
where $N_i =
N_i\left(\rho(\vect{r}),\vect{u}(\vect{r})\right)$ is a
polynomial function of the local density and velocity, which may be
found by discretizing the well-known Maxwell-Boltzmann equilibrium
distribution.
Our implementation uses the Shan-Chen
approach~\cite{bib:shan-chen}, by incorporating an explicit forcing term
in the collision operator in order to model multicomponent interacting
fluids. Shan and Chen extended $f_i$ to the form $f_i^\sigma$, where each component is denoted by a
different value of the superscript $\sigma$, so that density and
momentum of a component $\sigma$ are given by $\rhosig = \sumi
\fis$ and $\rhosig \vect{u}^\sigma = \sumi \fis \ci$. The
fluid viscosity $\nu^\sigma$ is proportional to $(\tau^\sigma-1/2)$ and
the particle mass is $m^\sigma$. This results in a lattice BGK
equation (\ref{eq:lbgk}) of the form 
\begin{equation} \label{eq:lbgk-sc}
\fis(\vect{r},t+1) - \fis(\vect{r},t) = - \frac 1 \tausig \left[ \fis -
N_i(\rhosig, \vect{v}^\sigma) \right] 
\end{equation} 
The velocity
$\vect{v}^\sigma$ is found by calculating a weighted average velocity
$\vect{u}'$ and then adding a term to account for external forces:
\begin{equation} 
\vect{u}' = \left( \sumsig \frac \rhosig \tausig
\vect{u}^\sigma \right) / \left( \sumsig \frac \rhosig \tausig \right),
\qquad \vect{v}^\sigma = \vect{u}' + \frac \tausig \rhosig
\vect{F}^\sigma . 
\end{equation} 
In order to produce nearest-neighbour interactions between components, the
force term assumes the form 
\begin{equation}
\label{eq:colour-colour} \vect{F}^\sigma = - \psisig ( \vect{x} )
\sumsigb g_{\sigma \bar{\sigma}} \sumi \psisigb \left( \vect{x} + \ci
\right) \ci, 
\end{equation} 
where $\psisig ( \vect{x} ) = \psisig (
\rhosig ( \vect{x}))$ is an effective charge for component $\sigma$;
$g_{\sigma \bar{\sigma}}$ is a coupling constant controlling the
strength of the interaction between two components $\sigma$ and
$\bar{\sigma}$. If $g_{\sigma \bar{\sigma}}$ is set to zero for $\sigma
= \bar{\sigma}$, and a positive value for $\sigma \neq \bar{\sigma}$
then, in the interface between bulk regions of each component, particles
experience a force in the direction away from the interface, producing
immiscibility. In two-component systems, it is usually the case that
$g_{\sigma \bar{\sigma}} = g_{\bar{\sigma}\sigma} = g_{br}$.
Amongst other things, this model has been used to simulate spinodal
decomposition~\cite{bib:chin-coveney,bib:gonzalez-nekovee-coveney},
polymer blends~\cite{bib:martys-douglas}, liquid-gas phase
transitions~\cite{bib:shan-chen-liq-gas}, and flow in porous
media~\cite{bib:martys-chen}.
Amphiphilic fluids may be
treated by introducing a new species of particle
with an orientational degree of
freedom, which is modelled by a vector dipole moment
$\vect{d}$~\cite{bib:chen-boghosian-coveney} with magnitude $d_0$.
The dipole field
$\vect{d}(\vect{x},t)$ represents the average orientation of any
amphiphile present at site $\vect{x}$. During advection, values of
$\vect{d}(\vect{x},t)$ are propagated according to (tildes denote
post-collision values)
\begin{equation}
\rho^{\mathrm s}(\vect{x},t+1) \vect{d}(\vect{x},t+1)
= \sumi \tilde{f_i^{\mathrm s}}(\vect{x}-\ci,t)
	\tilde{\vect{d}}(\vect{x}-\ci,t),
\end{equation}
During collision, the dipole moments evolve in a BGK process
controlled by a dipole relaxation time $\tau_d$:
\begin{equation}
\tilde{\vect{d}}(\vect{x},t) = \vect{d}(\vect{x},t)
- \frac 1 {\tau_d}
\left[
	\vect{d}(\vect{x},t)-\vect{d}^{\mathrm{(eq)}}(\vect{x},t)
\right].
\end{equation}
The equilibrium dipole moment $\vect{d}^{\mathrm{(eq)}}\simeq {\beta
d_0}\vect{h} / 3$ is
aligned with the colour field $\vect{h}$
which contains a component $\vect{h}^c$
due to coloured particles, and a part $\vect{h}^s$
due to dipoles. With $q^\sigma$ being a colour charge, such as $+1$ for
red
particles, $-1$ for blue particles, and $0$ for amphiphile particles,
one gets
\begin{equation}
\vect{h}^c = \sumsig q^\sigma
	\sumi \rhosig(\vect{x}+\ci) \ci,
\end{equation}
\begin{equation}
\vect{h}^s (\vect{x},t) = \sumi \left[
\sum_{j \neq 0} f_i^s ( \vect{x} + \ci, t) 
	\vect{\theta}_j \cdot \vect{d}_i (\vect{x}+\vect{c}_j,t)
	+ f_i^s (\vect{x},t) \vect{d}_i(\vect{x},t)
\right].
\end{equation}
The second-rank tensor $\vect{\theta}_j$ is defined in
terms of the unit tensor $\vect{I}$ and lattice vector $\vect{c}_j$ as
$\vect{\theta}_j = \vect{I} -  D \vect{c}_j \vect{c}_j / {c^2}$.
In the presence of an amphiphilic species, the force on a coloured
particle includes an additional term $\vect{F}^{\sigma,s}$ to account
for the colour field due to the amphiphiles. By treating an amphiphilic
particle as a pair of oil and water particles with a very small
separation $\vect{d}$, introducing a constant $g_{\sigma s}$ to control
the strength of
the interaction between amphiphiles and non-amphiphiles and Taylor-expanding in $\vect{d}$, it can be
shown that this term is given by 
\begin{equation}
\vect{F}^{\sigma,s} \xt  = 
	-2 \psisig\xt g_{\sigma s}
	\sum_{i \neq 0}
		\tilde{\vect{d}}\xpct \cdot \vect{\theta}_i
			\psi^{s} \xpct.
\end{equation}
While amphiphiles do not possess a net colour charge, they also
experience a force due to the colour field, consisting of a part
$\vect{F}^{s,c}$ due to ordinary species, and a part $\vect{F}^{s,s}$
due to other amphiphiles:
\begin{equation}
\vect{F}^{s,c} = 
	2 \psi^{s} \xt \tilde{\vect{d}}\xt \cdot
	\sumsig g_{\sigma s}
	\sum_{i \neq 0} \vect{\theta}_i \psisig \xpct,
\end{equation}
\begin{eqnarray}
\vect{F}^{s,s} &=& 
	- \frac {4D} {c^2}
	g_{ss} \psi^{s}(\vect{x})
	\sumi \left\{ 
		\tilde{\vect{d}}\xpc \cdot \vect{\theta}_i
		\cdot \tilde{\vect{d}}(\vect{x}) \ci
		\right.
		\\
		\nonumber
		&+& \left. \left[
			\tilde{\vect{d}}\xpc
			\tilde{\vect{d}}(\vect{x})
			+ \tilde{\vect{d}}(\vect{x})
			  \tilde{\vect{d}}\xpc
		\right] 
		\cdot \ci
	\right\}
	\psi^{s} \xpc . 
	\nonumber
\end{eqnarray}
While the form of the interactions seems straightforward at a mesoscopic
level, it is essentially phenomenological, and it is not necessarily easy
to relate the interaction scheme or its coupling constants to either
microscopic molecular characteristics, or to macroscopic phase behaviour.
The phase behaviour can be very difficult to predict beforehand from the
simulation parameters, and brute-force parameter searches are often
resorted to~\cite{bib:boghosian-coveney-love}.

\section{Technical projects}
Our three-dimensional lattice Boltzmann code, LB3D, is written in
Fortran 90 and designed to run on distributed-memory parallel computers,
using MPI for communication. In each simulation, the fluid is
discretized onto a cuboidal lattice, each lattice point containing
information about the fluid in the corresponding region of space. Each
lattice site requires about a kilobyte of memory per lattice site so
that, for example, a simulation on a $128^3$ lattice would require
around $2.2{\mathrm{GB}}$ memory.
The high-performance computing machines on which most of the simulation
work is performed are typically rather heavily used
The situation frequently arises that while a simulation is running on one machine, CPU time
becomes available on another machine which may be able to run the job
faster or cheaper. The LB3D program has the ability to ``checkpoint''
its entire state to a file. This file can then be moved to another
machine, and the simulation restarted there, even if the new machine has
a different number of CPUs or even a completely different architecture.
It has been verified that the simulation results are independent of the
machine on which the calculation runs, so that a single simulation may
be migrated between different machines as necessary without affecting
its output.
As a conservative rule of thumb, the code runs at over $10^4$ lattice site
updates per second per CPU on a fairly recent machine, and has been
observed to have roughly linear scaling up to order $10^3$ compute nodes.
A $128^3$ simulation contains around $2.1 \times 10^6$ lattice sites;
running it for 1000 timesteps requires  about an hour of real time, split
across $64$ CPUs. The largest simulation we performed used a $1024^3$
lattice.
The output from a simulation usually takes the form of a single
floating-point number for each lattice site, representing, for example,
the density of a particular fluid component at that site. Therefore, a
density field snapshot from a $128^3$ system would produce output files of
around $8{\mathrm{MB}}$. Writing data to disk is one of the bottlenecks in
large scale simulations. If one simulates a 1024$^3$ system, each data
file is $4{\mathrm{GB}}$ in size. LB3D is able to benefit from the
parallel filesystems available on many large machines today, by using the
MPI-IO based parallel HDF5 data format \cite{bib:hdf5}.
Our code is very robust regarding different platforms or cluster
interconnects: even with moderate inter-node bandwidths it achieves
almost linear scaling for large processor counts with the only
limitation being the available memory per node. The platforms our code
has been successfully used on include various supercomputers like the
IBM pSeries, SGI Altix and Origin, Cray T3E, Compaq Alpha clusters, NEC
SX6, as well as low cost 32- and 64-bit Linux clusters.
However, due to compiler or machine peculiarities it is a time consuming
task to achieve optimum performance on many different platforms. Porting a
complex Fortran code like LB3D to new platforms is often very difficult
and time-consuming without the assistance of well trained staff at the
corresponding computer centres. Some of these problems are due to
portability issues with the Fortran language. Also, tuning a code to take
full advantage of the machine on which it runs requires considerable
knowledge of the local system's quirks. It is hoped that some of the
portability issues could be solved in future by well-designed middleware.
Such issues include the fact that location, size, and duration of
temporary filespace change from machine to machine, as do the methods for
invoking compilers and batch queues.

LB3D has successfully been used to study various problems like spinodal
decomposition with and without shear
\cite{bib:gonzalez-nekovee-coveney,bib:harting-venturoli-coveney}, flow in
porous media \cite{bib:harting-venturoli-coveney}, the self-assembly of
cubic mesophases such as the 'P'-phase \cite{Maziar:2001} in binary
water-surfactant systems, or the cubic gyroid phase in ternary amphiphilic
systems \cite{bib:gonzalez-coveney,bib:gonzalez-coveney-2}.  Before we
were able to take advantage of computational steering techniques, our work
usually involved large scale parameter searches organised as taskfarming
jobs in order to find the areas of interest of the available parameter
space. 
The technique of computational
steering\cite{bib:chin-harting-jha,bib:brooke-coveney-harting,bib:love-nekovee-coveney-chin-gonzalez-martin}
has been used successfully in smaller-scale simulations to optimize
resource usage. Typically, the procedure for running a simulation of the
self-assembly of a mesophase would be to set up the initial conditions,
and then submit a batch job to run for a certain, fixed number of
timesteps. If the timescale for structural assembly is unknown then the
initial number of timesteps for which the simulation runs is, at best,
an educated guess. It is not uncommon to examine the results of such a
simulation once they return from the batch queue, only to find that a
simulation has not been run for sufficient time (in which case it must
be tediously resubmitted), or that it ran for too long, and the majority
of the computer time was wasted on simulation of an uninteresting
equilibrium system showing no dynamical behaviour.
Another unfortunate scenario often occurs when the phase diagram of a
simulated system is not well known, in which case a simulation may
evolve away from a situation of interest, wasting further CPU time.
Computational steering, the ability to watch and control a calculation
as it runs, can be used to avoid these difficulties: a simulation which
has equilibrated may be spotted and terminated, preventing wastage of
CPU time.
More powerfully, a simulation may be steered through parameter space
until it is unambiguously seen to be producing interesting results: this
technique is very powerful when searching for emergent phenomena, such
as the formation of surfactant micelles, which are not clearly related
to the underlying simulation parameters.
Steering is performed using the RealityGrid steering library which has
been developed by collaborators at the University of Manchester. The
library was built with the intention of making it possible to add
steering capabilities to existing simulation codes with as few changes
as possible, and in as general a manner as possible. Once the
application has initialized the steering library and informed it which
parameters are to be steered, then after every timestep of the
simulation, it is possible to perform tasks such as checkpointing the
simulation, saving output data, stopping the simulation, or restarting
from an existing checkpoint.
When a steered simulation is started, a Steering Grid Service (SGS) is
also created, to represent the steerable simulation on the Grid. The SGS
publishes its location to a Registry service, so that steering clients
may find it. This design means that it is possible for clients to
dynamically attach to and detach from running simulations.

Successful computational steering requires that the simulation operators
have a good understanding of what the simulation is doing, in real time:
this in turn requires good visualization capabilities. Each running
simulation emits output files after certain periods of simulation time
have elapsed. The period between output emission is initially determined
by guessing a timescale over which the simulation will change in a
substantial way; however, this period is a steerable parameter, so that
the output rate can be adjusted for optimum visualization without
producing an excessive amount of data.
The LB3D code itself will only emit volumetric datasets as described
above; these must then be rendered into a human-comprehensible form
through techniques including volume-rendering, isosurfacing,
ray-tracing, slice planes, and Fourier transforms. The process of
producing such comprehensible data from the raw datasets is itself
computationally intensive, particularly if it is to be performed in real
time, as required for computational steering.
For this reason, we use separate visualization clusters to render the
data. Output volumes are sent from the
simulation machine to the remote visualization machine, so that the
simulation can proceed independently of the visualization; these are
then rendered using the open source VTK\cite{bib:vtk} visualization
library into bitmap images, which can in turn be multicast over the
AccessGrid, so that the state of the
simulation can be viewed by scientists around the globe. In particular,
this was demonstrated by performing and interacting with a simulation in
front of a live worldwide audience, as part of the SCGlobal track of the
SuperComputing 2004 conference.
The RealityGrid steering architecture was designed in a sufficiently
general manner that visualization services can also be represented by
Steering Grid Services: in order to establish a connection between the
visualization process and the corresponding simulation, the simulation
SGS can be found through the Registry, and then interrogated for the
information required to open the link.

In order to be able to deploy the above described components as part of
a usable simulation Grid, a substantial amount of coordination is
necessary, so that the end user is able to launch an entire simulation
pipeline, containing migratable simulation, visualization, and steering
components, from a unified interface. This requires a system for keeping
track of which services are available, which components are running,
taking care of the checkpoints and data which are generated, and to
harmonize communication between the different components.
This was achieved through the development of a Registry service,
implemented using the {\tt{OGSI:~\!\!\!\!:Lite}} \cite{bib:ogsilite}
toolkit. The RealityGrid steering library\cite{bib:chin-harting-jha}
communicates with the rest of the Grid by exposing itself as a
``Grid Service''. Through the Registry service, steering clients
are able to find, dynamically attach to, communicate with, and detach from
steering services to control a simulation or visualization process.

Large lattices require a highly scalable code, access to high performance
computing, terascale storage facilities and high performance
visualisation. LB3D provides the first of these, while the others are
being delivered by the major computing centres.  
We expect to be able to run our simulations in an even more efficient
way due to the significant worldwide effort being invested in the
development of reliable computational grids. These are a collection of
geographically distributed and dynamically varying resources, each
providing services such as compute cycles, visualization, storage, or
even experimental facilities. The major difference between computational
grids and traditional distributed computing is the transparent sharing
and collective use of resources, which would otherwise be individual and
isolated facilities. Perhaps at some point computational grids will
offer information technology what electricity grids offer for other
aspects of our daily life: a transparent and reliable resource that is
easy to use and conforms to commonly agreed standards
\cite{gridbook2,bib:Berman}. Robust and smart middleware will find the
best available resources in a transparent way without the user having to
care about their location.
Unfortunately, reliable and robust computational grids are not available
yet. We used various different demonstration grids which were assembled
especially for a given event or were intended for use as prototyping
platforms rather than usable production grids.
These mainly included grids
coupling major compute resources in the UK and the biggest effort took
place within the TeraGyroid project
\cite{bib:teragyroid,bib:teragyroid-epsrc} where the main
machines of the UK's national HPC centres were coupled with the TeraGrid
facilities in the US through a custom high-performance
network. In total, about 5000 CPUs were part of this grid.
Collaborative steering sessions with active participants on two
continents and observers worldwide were made possible through this
approach.

\section{Scientific projects}
\subsection{Complex fluids under shear}
\label{Sec:Shear}
\begin{figure}[h]
\begin{center}
\includegraphics[height=4cm]{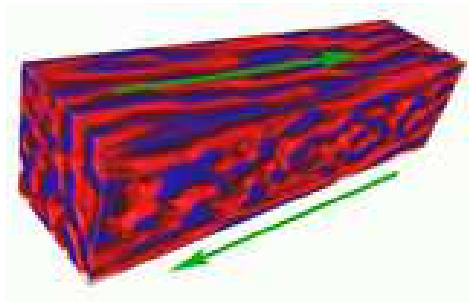}
\end{center}
\caption{Spinodal decomposition under shear. Differently coloured regions
denote the majority of the corresponding fluid. The arrows depict the
movement of the sheared boundaries (movie available in online version).
}
\label{fig:spinodal-shear}
\end{figure}

In many industrial applications, complex fluids are subject to shear
forces. For example, axial bearings are often filled with fluid to
reduce friction and transport heat away from the most vulnerable parts
of the device. It is very important to understand how these fluids
behave under high shear forces, in order to be able to build reliable
machines and choose the proper fluid for different applications.
In our simulations we use Lees-Edwards boundary conditions, which were
originally developed for molecular dynamics simulations in 1972
\cite{bib:lees-edwards} and have been used in lattice Boltzmann
simulations by different authors before
\cite{bib:wagner-yeomans-shear,bib:wagner-pagonabarraga,bib:harting-venturoli-coveney}. 
We applied our model to study the behaviour of binary immiscible and
ternary amphiphilic fluids under constant and oscillatory shear. In the
case of spinodal decomposition under constant shear, the first results
have been published in \cite{bib:harting-venturoli-coveney}. The phase
separation of binary immiscible fluids without shear has been studied in
detail by different authors, and LB3D has been shown to model the
underlying physics successfully \cite{bib:gonzalez-nekovee-coveney}. In
the non-sheared studies of spinodal decomposition it has been shown that
lattice sizes need to be large in order to overcome finite size effects:
 128$^3$ was the minimum acceptable number of lattice
sites \cite{bib:gonzalez-nekovee-coveney}. For high shear rates, systems
also have to be very long because, if the system is too small, the
domains interconnect across the $\bf z = 0$ and $\bf z = nz$ boundaries
to form interconnected lamellae in the direction of shear. Such
artefacts need to be eliminated from our simulations. Figure
\ref{fig:spinodal-shear} shows an example from a simulation with lattice
size 128x128x512. The volume rendered blue and red areas depict the
different fluid species and the arrows denote the direction of shear.
In the case of ternary amphiphilic fluid mixtures under shear we are
interested in the influence of the presence of surfactant molecules on the
phase separation. We also study the stress response and stability of
cubic mesophases such as the gyroid phase \cite{bib:gonzalez-coveney} or
the ``P''-phase \cite{bib:nekovee-coveney} under shear. Such complex fluids
are expected to exhibit non-Newtonian properties (see below).
Computational steering has turned out to be very useful for checking on
finite size effects during a sheared fluid simulation, since the human
eye is extremely good at spotting the sort of structures indicative of
such effects. Implementing an algorithm to automatically
recognize ``unphysical'' behaviour is a highly nontrivial task in
comparison.

\subsection{Flow in porous media} \label{Sec:Porous}
Studying transport phenomena in porous media is of great interest in
fields ranging from oil recovery and water purification to industrial
processes like catalysis. In particular, the oilfield industry uses
complex, non-Newtonian, multicomponent fluids (containing polymers,
surfactants and/or colloids, brine, oil and/or gas), for processes like
fracturing, well stimulation and enhanced oil recovery. The rheology and
flow behaviour of these complex fluids in a rock is different from their
bulk properties. It is therefore of considerable interest to be able to
characterise and predict the flow of these fluids in porous media.
From the point of view of a modelling approach, the treatment of complex
fluids in three-dimensional complex geometries is an ambitious goal since
the lattice has to be large enough to resolve individual structures. The
advantage of lattice Boltzmann (or lattice gas) techniques is that
complex geometries can be modelled with ease.
\begin{figure}
\centerline{\includegraphics[height=4cm]{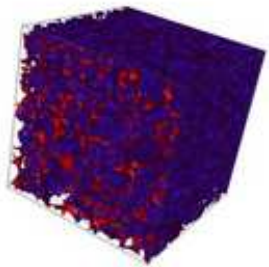}}
\caption{
Rendering of $4.9\mu m$ resolution X-ray microtomographic data of a 
$512^3$ sample of Bentheimer sandstone.
The pore space is shown in red, while the rock is represented in blue.}
\label{f:Benth}
\end{figure}
Synchrotron based X-ray microtomography (XMT) imaging techniques provide
high resolution, three-dimensional digitised images of rock samples. By
using the lattice Boltzmann approach in combination with these high
resolution images of rocks, not only is it possible to compute
macroscopic transport coefficients, such as the permeability of the
medium, but information on local fields, such as velocity or fluid
densities, can also be obtained at the pore scale, providing a detailed
insight into local flow characterisation and supporting the
interpretation of experimental measurements \cite{bib:auzeraisGRL96}.
The XMT technique measures the linear attenuation coefficient from which
the mineral concentration and composition of the rock can be computed.
Morphological properties of the void space, such as pore size distribution
and tortuosity, can be derived from the tomographic image of the rock
volume, and the permeability and conductivity of the rock can be
computed \cite{bib:spannePRL96}. The tomographic data are represented by
a reflectivity greyscale value, where the linear size of each voxel is
defined by the imaging resolution, which is usually on the order of
microns. By introducing a threshold to discriminate between pore sites
and rock sites, these images can be reduced to a binary (0's and 1's)
representation of the rock geometry. Utilizing the lattice Boltzmann
method, single phase or multiphase flow can then be described in these
real porous media.

Lattice Boltzmann and lattice gas techniques have already been applied to
study single and multiphase flow through three-dimensional
microtomographic reconstruction of porous media. For example, Martys and
Chen \cite{bib:martys-chen} and Ferr{\'e}ol and Rothman
\cite{bib:ferreol-rothman} studied relative permeabilities of binary
mixtures in Fontainebleau sandstone. These studies validated the model
and the simulation techniques, but were limited to small lattice sizes, of
the order of $64^3$.
Simulating fluid flow in real rock samples allows us to compare
simulation data with experimental results obtained on the same, or
similar, pieces of rock. For a reasonable comparison, the size of the
rock used in lattice Boltzmann simulations should be of the same order
of magnitude as the system used in the experiments, or at least large
enough to capture the rock's topological features. The more
inhomogeneous the rock, the larger the sample size needs to be in order
to describe the correct pore distribution and connectivity. 
Another reason for needing to use large
lattice sizes is the influence of boundary conditions and lattice
resolution on the accuracy of the lattice Boltzmann method. It has been
shown (see for example \cite{bib:He97}, \cite{bib:chen-doolen} and
references therein) that the Bhatnagar-Gross-Krook (BGK) \cite{bib:bgk}
approximation of the lattice Boltzmann equation which is commonly used
causes so-called bounce-back boundaries to become inaccurate, resulting in
effects such as the computed permeability being a function of the viscosity.
This effect can be limited by lowering the viscosity and increasing the
lattice resolution.
To accurately describe hydrodynamic behaviour using lattice Boltzmann
simulations, the Knudsen number, which represents the ratio of the mean free
path of the fluid particles and the characteristic length scale of the
system (such as the pore diameter), has to be small. If the pores are
resolved with an insufficient number of lattice points, finite size effects
arise, leading to an inaccurate description of the flow field. In
practice, at least five to ten lattice sites are needed to resolve a
single pore. Therefore, in order
to be able to simulate realistic sample sizes, we need large lattices of
the order of 512$^3$.

\begin{figure}
\centerline{\includegraphics[width=10cm]{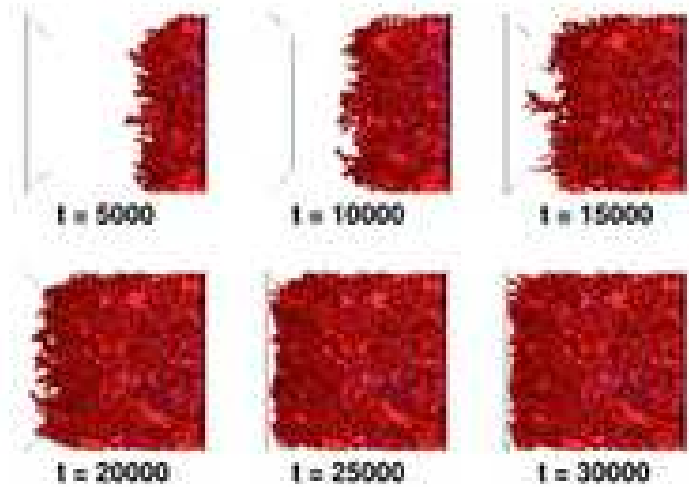}}
\caption{An originally fully fluid saturated rock is being invaded by
another immiscible fluid using a body force $g_{\rm accn}$ = 0.003. The
oil slowly pushes the other fluid component out of the rock pores
until the rock is fully saturated by oil at $t$ = 30000. For
better visability only the invading fluid is shown (movie available in
online version).}
\label{fig:invasion}
\end{figure}

Using LB3D, we are able to simulate drainage and imbibition processes in
a $512^3$ subsample of Bentheimer sandstone X-ray tomographic data. The
whole set of XMT data represented the image of a Bentheimer sample of
cylindrical shape with diameter 4mm and length 3mm. The XMT data were
obtained at the European Synchrotron Research Facility (Grenoble) at a
resolution of $4.9 \mu {\rm m}$, resulting in a data set of
approximately 816x816x612 voxels. Figure \ref{f:Benth} shows a snapshot
of the $512^3$ subsystem.
We compare simulated velocity distributions with experimentally obtained
magnetic resonance imaging (MRI) data of oil and brine infiltration into
saturated Bentheimer rock core \cite{MRISheppard}. The rock sample used in
these MRI experiments had a diameter of 38 mm and was 70 mm long and was
imaged with a resolution of 280 microns. The system simulated was smaller,
but still of a similar order of magnitude and large enough to represent
the rock geometry. On the other hand, the higher space resolution provided
by the simulations allows a detailed characterisation of the flow field in
the pore space, hence providing a useful tool to interpret the MRI
experiments, for example in identifying regions of stagnant fluid.
Figure \ref{fig:invasion} shows an example from a binary invasion study.
A rock which is initially fully saturated with ``water'' (blue), is
being invaded by ``oil'' (red) from the right side. The lattice size is
$512^3$ and the forcing level is set to $g_{\rm accn}$ = 0.003. In
figure \ref{fig:invasion}, only the invading fluid component is shown,
i.e. only areas where oil is the majority component are rendered.
Periodic boundary conditions are applied, and fluid leaving the system
on the left side is converted to oil before re-entering on the opposite
side. After 5000 timesteps, the oil has invaded about one quarter of the
system already and after 25000 timesteps only small regions of the rock
pore space are still filled with water. After 30000 timesteps, the water
component has been fully pushed out of the rock.
This example only covers binary (oil/water) mixtures of Newtonian fluids,
since this is a first and necessary step in the understanding of
multiphase fluid flow in porous media
\cite{bib:harting-venturoli-coveney}. However, we are able to study the
flow of binary immiscible fluids with an additional amphiphilic component
in porous media and expect results to be presented elsewhere in the near
future.

\subsection{The cubic gyroid mesophase}
\begin{figure}[h]
\begin{center}
\includegraphics[height=4cm]{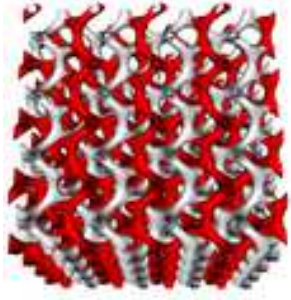}
\end{center}
\caption{Structure of the two labyrinths
enclosed by a gyroid minimal surface, showing the characteristic triple
junctions.}
\label{fig:gyroid-wishbones}
\end{figure}

It was recently shown by Gonz\'{a}lez and
Coveney\cite{bib:gonzalez-coveney} that the dynamical self-assembly of a
particular amphiphile mesophase, the gyroid, can be modelled using the
lattice Boltzmann method. This mesophase was observed to form from a
homogeneous mixture, without any external constraints imposed to bring
about the gyroid geometry, which is an emergent effect of the mesoscopic
fluid parameters.
It is important to note that this method allows examination of the dynamics of
mesophase formation, since most treatments to date have focussed on properties
or mathematical
description\cite{bib:seddon-templer-2,bib:schwarz-gompper-2,bib:gandy-klinowski,bib:grosse-brauckmann}
of the static equilibrium state. In addition to its biological importance,
there have been recent attempts\cite{bib:chan-hoffman-etal} to use
self-assembling gyroids to construct nanoporous materials.
During the gyroid self-assembly process, several small, separated
gyroid-phase regions or domains may start to form, and then grow. Since
the domains evolve independently, the independent gyroid regions will
in general not be identical, and can differ in orientation, position, or
unit cell size; grain-boundary defects arise between gyroid domains.
Inside a domain, there may be dislocations, or line defects,
corresponding to the termination of a plane of unit cells; there may
also be localised non-gyroid regions, corresponding to defects due to
contamination or inhomogeneities in the initial conditions.
Understanding such defects is therefore important for our knowledge of
the dynamics of surfactant systems, and crucial for an understanding of
how best to produce mesophases experimentally and industrially.
\begin{figure}[h]
\begin{center}
\includegraphics[height=4cm]{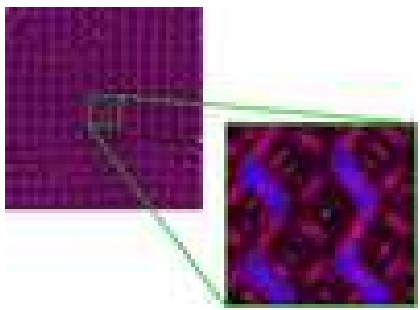}
\end{center}
\caption{A volume rendered dataset of a 128$^3$ system after 100000
simulation timesteps. Various gyroid domains have formed and the
close-up shows the extremely regular, crystalline, gyroid structure within
a domain (movies available in online version).}
\label{fig:gyroid}
\end{figure}
In small-scale simulations of the gyroid, the mesophase will evolve to
fill the simulated region perfectly, without defects. As the lattice size
grows, it becomes more probable that multiple gyroid domains will emerge
independently, so that grain boundary defects are more likely to appear,
and the time required for localized defects to diffuse across the lattice
increases, making it more likely that defects will persist. Therefore,
examination of the defect behaviour of surfactant mesophases requires the
simulation of very large systems.
Figure \ref{fig:gyroid} shows an example of a 128$^3$ system after 100000
simulation timesteps. Multiple gyroid domains have formed and the close-up
shows the extremely regular, crystalline, gyroid structure within a
domain. Figure \ref{fig:gyroid-wishbones} demonstrates some of the most
interesting properties of the gyroid mesophase: two labyrinths mainly
consisting of water and oil counterparts are enclosed by the gyroid
minimal surface at which the surfactant molecules accumulate. The
characteristic triple junctions can be seen clearly.

The TeraGyroid experiment\cite{bib:teragyroid,bib:teragyroid-epsrc} addressed a large scale
problem of genuine scientific interest and showed how intercontinental
grids permit the use of novel techniques in collaborative computational
science, which can dramatically reduce the time to insight. TeraGyroid
used computational steering over a Grid to study the self-assembly and
dynamics of gyroid mesophases using the largest set of lattice Boltzmann
simulations ever performed. Around the Supercomputing 2003 conference we
were able to simulate gyroid formation and defect behaviour harnessing
the compute power of a large fraction of the UK and US HPC facilities.
Altogether we were able to use about 400000 CPU hours and generate two
terabytes of simulation data.
\begin{figure}[h]
\begin{center}
\includegraphics[height=4cm]{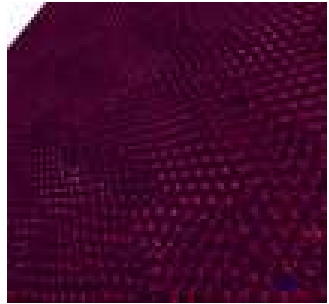}
\end{center}
\caption{In order to eliminate finite size effects from simulations, very
large lattice sizes are needed. If one is interested in the statistical
behaviour of defects, then the lattice size has to be increased even more,
since otherwise only a limited number of defects can be found in the
system.  This figure shows a snapshot from what we believe to be the
largest ternary lattice Boltzmann simulation ever performed, on a 1024$^3$
lattice. \label{fig:gyroid1024} }
\end{figure}
In order to make sure our simulations are virtually free of finite size
effects, we simulated different system sizes from 64$^3$ to 1024$^3$,
usually for about 100000 timesteps. In order to study the long term
behaviour of the gyroid mesophase, some simulations have even run for one
million timesteps. For 100000 timesteps we found that 256$^3$ or even
128$^3$ simulations do not suffer from finite size effects, but after very
long simulation times we might even have to move to larger lattices.
Even with the longest possible simulation times, we were not able to
generate a ``perfect'' crystal. Instead, either differently orientated
domains can still be found or individual defects are still moving
around. It is of particular interest to study the exact behaviour of
the defect movement, which can be done by gathering statistics of the
simulation data by counting and tracking individual defects. Gathering
useful statistics implies large numbers of measurements and therefore
large lattices, which is the reason for the 512$^3$ and 1024$^3$
simulations performed. The memory requirements exceed the available
resources on most supercomputers and limits us to a small number of
machines. Also, it requires substantial amounts of CPU time to reach
suffcient simulation times. In the case of the 1024$^3$ system, 2048
CPUs of a recent Compaq Alpha cluster are only able to simulate about
100 simulation timesteps per hour. Running for 100000 timesteps would
require more than two million CPU hours or 42 days and is therefore
unfeasible. Also, handling the data files which are $4{\mathrm{GB}}$
each and checkpoint files which are $0.5{\mathrm{TB}}$ each is very
awkward with the infrastructure available today. In order to be able to
gain useful data from the large simulations, we first run a 128$^3$
system with periodic boundary conditions, until it forms a gyroid. This
system is then duplicated 512 times to produce a 1024$^3$ gyroid system.
In order to reduce effects introduced due to the periodic upscaling, we
perturb the system and let it evolve. We anticipate that the unphysical
effects introduced by the upscaling process will decay after a
comparably small number of timesteps, thus resulting in a system that is
comparable to one that started from a random mixture of fluids. This has
to be justified by comparison with data obtained from test runs performed
on smaller systems. Figure \ref{fig:gyroid1024} shows a snapshot of a
volume rendered dataset from the upscaled 1024$^3$ system at 1000
timesteps after the upscaling process. The unphysical periodic
structures introduced by the individual 128$^3$ systems can still
clearly be seen. 

\begin{figure}[h]
\begin{center}
\includegraphics[height=3cm]{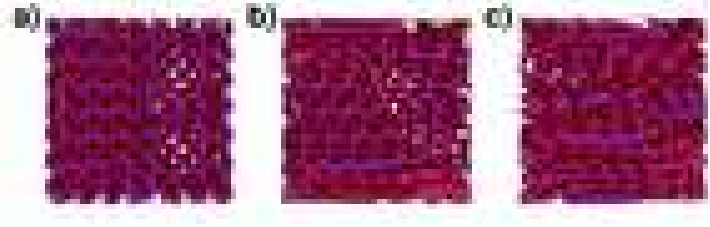}
\end{center}
\caption{a sheared gyroid mesophase: a) before the onset of shear, b)
at the onset of shear, c) after long shear times.
}
\label{fig:gyroid-shear}
\end{figure}
Currently, work is in progress to study the stability of the gyroid
mesophase. We are interested in the influence of perturbation on a
gyroid and the strength of the perturbation needed to break up a well
developed mesophase. Similar studies are performed experimentally by
applying constant or oscillatory shear. Here, we study the dependence of
the gyroid stability on the shear rate, and expect to find evidence of
the non-Newtonian properties of the fluid. An example from those studies
can be seen in figure \ref{fig:gyroid-shear}, which shows three
snapshots of the same simulation. The first shows the liquid crystal
before the onset of shear, the second only a few hundred timesteps after
shear has been turned on and the third image demonstrates how the gyroid
melts if the shear stress becomes too strong.

\begin{figure}[h]
\begin{center}
\includegraphics[height=5cm]{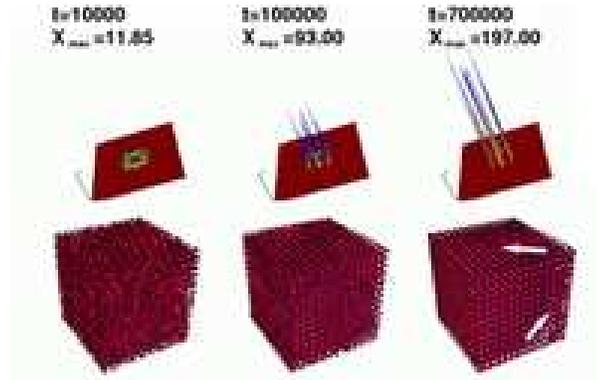}
\end{center}
\caption{Three-dimensional structure factor of the order parameter at
timesteps $t$=10000, 100000, and 700000, lattice size 128$^3$ and
simulation parameters as given in the text. 
For comparability with SAXS experimental data, we display
the total structure factor in the $x$-direction $X$=$\sum_{k_x}
S(\mathbf{k},t)$. $X_{max}$ denotes the value of the largest peak divided
by the number of lattice sites in the direction of summation (128 in this
case). The lower half of the figure shows volume rendered visualizations
of the corresponding order parameters and the white arrows are a guide for
the eye to spot some defective areas at the top surface and the right
corner at $t$=700000.}
\label{fig:3DFFT}
\end{figure}

\begin{figure}[h]
\begin{center}
\includegraphics[height=4cm]{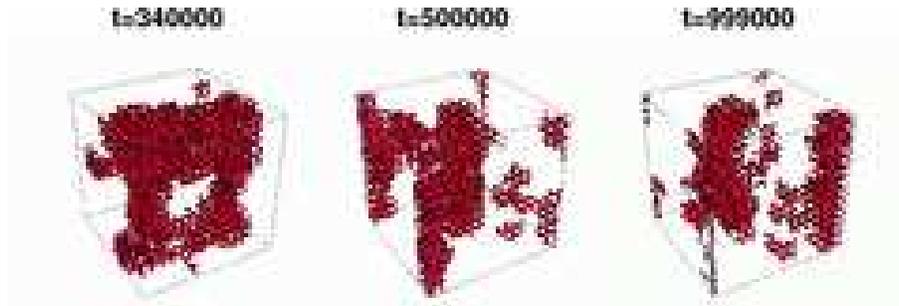}
\end{center}
\caption{Volume rendered visualization of the order parameter at t=340000,
500000, 999000. Only the defects are shown
as they have been isolated from the full datasets using the pattern
recognition algorithm (movie available in onine version).}
\label{fig:strucmaskreconst}
\end{figure}

As seen before, simulation data from liquid crystal dynamics can be
visualized using isosurfacing or volume rendering techniques. The human
eye has a remarkable ability to easily distinguish between regions where
the crystal structure is well developed and areas where it is not.
However, manual analysis of large amounts of simulation data is not
feasible. In the case of the TeraGyroid project, about two terabytes of
data would have to be checked and catalogued manually. This task would
keep an individual busy for years. Therefore, computational methods for
defect detection and tracking are required. Developing algorithms to
detect and track defects is a non-trivial task, however, since defects can
occur within and between domains of varying shapes and sizes and over a
wide variety of length and time scales. 
A standard method to analyse simulation data is the calculation of the
three-dimensional structure function 
$S(\mathbf{k},t)\equiv\frac{1}{V}\left|\phi^\prime_\mathbf{k}(t)\right|^2$,
where $V$ is the number of cites of the lattice,
$\phi^\prime_\mathbf{k}(t)$ the Fourier transform of the fluctuations of
the order parameter $\phi^\prime\equiv\phi-\left<\phi\right>$, and
$\mathbf{k}$ is the wave vector
\cite{bib:gonzalez-nekovee-coveney,bib:gonzalez-coveney-2}.
$S(\mathbf{k},t)$
can easily be calculated, but only gives general information about the
crystal development \cite{bib:hajduk,bib:laurer,bib:gonzalez-coveney}. It
does not allow one to detect where the defects are located or how many
there are, nor does it furnish access to information about the number
of differently oriented gyroid domains.
$S(\mathbf{k},t)$ is given for a 128$^3$ system at timesteps $t$=10000, 100000,
and 700000 in figure \ref{fig:3DFFT}. We simulate for one million
timesteps -- more than an order of magnitude longer than any other LB3D
simulation performed before the TeraGyroid \cite{bib:TeraGyroidWWW}
project. The initial condition of the simulation is a random mixture with
maximum densities of 0.7 for the immiscible fluids and 0.6 for surfactant.
The coupling constant $g_{ss}$ is set to -0.0045 and the coupling between
surfactant and the other fluids is set to $g_{cs}$=-0.006. In
order to compare our data to experimentally obtained SAXS data
\cite{bib:hajduk}, we sum the structure factor in the $x$-direction;
$X_{max}$ denotes the value of the largest peak normalised by the number
of lattice sites in the direction of summation (128 in this case)
\cite{bib:defect-paper}. 
Gyroid assembly is evident due to the eight peaks of the structure factor
which become higher with ongoing simulation time. At $t$=700000,
$X_{max}$ reaches 197.00 and most of the previously existing domains have
merged into a single one. Only a few defects are left of which two can be
spotted visually at the right corner of the volume rendered visualisation
and the centre of the top surface (denoted by the white arrows).
The structure factor analysis does not provide any information about the
size, position or number of individual defects in the system. Therefore,
we developed more advanced algorithms for the detection and tracking of
defects.
As a first order approach, the data to be analyzed can be reduced by
cutting the three-dimensional data sets into slabs and projecting them
onto a two-dimensional plane. By using a raytracing algorithm for the
projection, we obtain regular patterns in areas where the gyroid is
perfectly developed and solid planes in defective areas. We developed
two algorithms which use the projection data to separate the defective
areas from the perfect crystal. The first approach is based on a
generic pattern recognition algorithm and should work with all liquid
crystals that form a regular pattern, while the second has been
developed with our particular problem in mind and is not known to work
with systems other than the gyroid mesophase. However, it is about an
order of magnitude faster and the general principles underlying it
should be applicable to different systems as well.
The first approach is based on the regularity or periodicity of patterns
and was developed by Chetverikov and Hanbury in 2001
\cite{bib:chetverikov} who applied it to patterns from the textile
industry. It is assumed that defect-free patterns are homogeneous and show
some periodicity. The algorithm searches for areas which are significantly
less regular (i.e. aperiodic) than the bulk of the dataset by computing
regularity features for a set of windows and identifying defects as
outliers. The regularity is quantified by computing the periodicity of the
normalised autocorrelation function in polar coordinates. In short, for
every window a regularity value is computed. If this value differs by more
than a defined threshold value from the median of all window regularity
values, the area is accordingly classified as a defect. For a more
detailed description of the algorithm see
\cite{bib:chetverikov,bib:chetverikov2,bib:defect-paper}.
The second approach encapsulates knowledge about the patterns produced by
regular and defect regions. As a consequence, it is an order of magnitude
faster than the pattern recognition code.
For each slab image, the algorithm creates a regular mesh in areas where
the gyroid structure is well developed, and an irregular mesh in
defective areas. The regions of regular mesh are discarded, leaving only
mesh that describes the perimeters of defect regions.  A flood-fill
algorithm is applied to these datasets to locate distinct defect
regions.
The output data of both detection algorithms for all two-dimensional
projections of a three-dimensional dataset can be used to reconstruct
three-dimensional volume data that only consists of defect regions.
Figure \ref{fig:strucmaskreconst} shows reconstructed datasets at
$t$=340000, 500000 and 999000 which have been detected using the pattern
recognition approach. However, the results obtained from the mesh
generator are similar. Even at $t$=340000 a very large region of
the system has not yet formed a well defined gyroid phase. 160000
timesteps later, the main defects are pillar shaped ones at the centre and
at the corners of the visualised systems. Due to the periodic boundary
conditions, the corner defects are connected and should be regarded as a
single one. As can be seen from the analysis at $t$=999000, defects in the
gyroid mesophase are very stable in size as well as in their position.
The pattern recognition algorithm is less efficient than mesh generation.
However, it is not limited to simulations of gyroid mesophases and more
robust with regrd to small fluctuations of the dataset. In the gyroid
case, it is more efficient to use the results from the mesh generator to
select a smaller number of datasets for post-processing using the pattern
recognition algorithm since the computational effort involved in the
pattern recognition can be substantial. For a more detailed description
of the algorithms see \cite{bib:defect-paper}. Currently, we are working
on more geometrically based algorithms to efficiently detect defects and
results will be published elsewhere in the near future.

\section{Conclusions}
During the last two years, we have worked on various scientific projects
using our lattice Boltzmann code LB3D. All of these projects reached the
limits of the HPC resources available to us today. However, without the
benefits obtained from software development within the RealityGrid
project, none of these projects would have been possible at all. These
improvements include the steering facilities, code optimizations, IO
optimizations as well as the platform independent checkpointing and
migration routines which have been contributed by various people within
the project. Without the lightweight Grid Service Container
{\tt{OGSI:~\!\!\!\!:Lite}} \cite{bib:ogsilite} projects like the
TeraGyroid experiment would not have been possible since existing
middleware toolkits such as Globus are rather heavyweight, requiring
substantial effort and local tuning on the part of systems administrators
to install and maintain. This effort cannot be expected from the average
scientist who is planning to use a computational
grid\cite{bib:lgpaper}.
The simulation pipeline requires
simulation, visualization, and storage facilities to be available
simultaneously, at times when their human operators can reasonably
expected to be around. This is often dealt with by manual reservation of
resources by systems administrators, but the ideal solution would
involve automated advance reservation and co-allocation procedures.
The most exciting project involving RealityGrid during the last two
years was the TeraGyroid experiment. Hundreds of individuals have worked
together to build a transcontinental grid not only as a demonstrator for
the grid techniques available today, but to perform a scientific
project. Since we would not have been able to gain as many new results
from the simulations performed during that period without the active use
of grid technologies, we have shown that the advent of computational
grids will be of great benefit for computational scientists.

\section*{Acknowledgements}
We would like to thank S.~Jha, M.~Harvey and G.~Giupponi (University
College London), A.R.~Porter, and S.M.~Pickles (University of Manchester), N.~Gonz\'{a}lez-Segredo (FOM Institute for Atomic and Molecular Physics), and
E.S.~Boek and J.~Crawshaw (Schlumberger Cambridge Research) for fruitful
discussions and E.~Breitmoser from the Edinburgh Parallel Computing Centre
for her contributions to our lattice Boltzmann code. 
We are grateful to the U.K. Engineering and Physical Sciences Research
Council (EPSRC) for funding much of this research through RealityGrid
grant GR/R67699 and to EPSRC and the National Science Foundation (NSF) for
funding the TeraGyroid project. 
This work was partially supported by the National Science Foundation under
NRAC grant MCA04N014 and PACI grant ASC030006P, and utilized computer
resources at the Pittsburgh Supercomputer Center, the National
Computational Science Alliance and the TeraGrid.
We acknowledge the European Synchrotron Radiation Facility for provision
of synchrotron radiation facilities and we would like to thank P.~Cloetens
for assistance in using beamline ID19, as well as  J.~Elliott
and G.~Davis of Queen Mary, University of London, for their
work in collecting the raw data and reconstructing the x-ray
microtomography data sets used in our Bentheimer sandstone images.

\bibliographystyle{astron-jh} 
\bibliography{main}

\end{document}